\documentclass[10pt,conference]{IEEEtran}
\usepackage{amsmath,amssymb}
\usepackage{graphicx}
\usepackage{mathrsfs}
\usepackage{tikz}
\usepackage{booktabs}
\usepackage{floatflt}
\usepackage{enumerate}
\usepackage{psfrag}	
\usepackage{array}
\usepackage{multirow,hhline}
\usepackage{exscale}
\usepackage{color}
\usepackage{colortbl}
\usepackage{epsfig}
\usepackage{cite}
\usepackage{amsthm}

\newcommand{\mat}[1]{\ensuremath{\mathbf{#1}}}
\renewcommand{\vec}[1]{\ensuremath{\mathbf{#1}}}

\newtheorem{theorem}{Theorem}

\begin{document}

\IEEEoverridecommandlockouts
\title{Interference Alignment and Neutralization in a Cognitive 3-User MAC-Interference Channel: Degrees of Freedom}
\author{
\IEEEauthorblockN{Anas Chaaban and Aydin Sezgin}
\thanks{%
The authors are with the Emmy-Noether Research Group on Wireless Networks, Institute TAIT, Ulm University, 89081 Ulm, Germany. Email: anas.chaaban@uni-ulm.de, aydin.sezgin@uni-ulm.de. This work is supported by the German Research Foundation, Deutsche
Forschungsgemeinschaft (DFG), Germany, under grant SE 1697/3.%
}
}

\maketitle


\begin{abstract}
A network consisting of a point-to-point (P2P) link and a multiple access channel (MAC) sharing the same medium is considered. The resulting interference network, with three transmitters and two receivers is studied from degrees of freedom (DoF) perspective, with and without cognition. Several cognition variants are examined. Namely, the setup is studied with (1) no cognitive transmitters, (2) a cognitive P2P transmitter, (3) one cognitive MAC transmitter, and (4) with two cognitive MAC transmitters. It is shown that having a cognitive P2P transmitter does not bring any DoF gain to the network. This is obtained by showing that the DoF of the two former cases (1) and (2) is 1. However, it is shown that a cognitive MAC transmitter is more beneficial since the latter two cases (3) and (4) have 3/2 DoF. The achievability of  3/2 DoF is guaranteed by using a combination of interference neutralization and interference alignment.
\end{abstract}


\section{Introduction}
Cognitive networks have witnessed increasing research attention recently. The idea of cognition was introduced to help achieve higher spectral efficiency in wireless networks and allow new communication systems to exist. A cognitive transmitter can co-exist with a primary system by establishing cooperation with it, in a way that boosts the primary system performance while sending a message of its own to its respective receiver. 

The most basic cognitive network is the cognitive interference channel (CIC) that was introduced in \cite{DevroyeMitranTarokh}. It is a setup where a cognitive point-to-point system communicates over the same medium as a primary point-to-point system. The cognitive transmitter knows the message of the primary transmitter non-causally. The capacity of this setup was obtained for several cases in \cite{WuVishwanathArapostathis,JovicicViswanath,MaricYatesKramer,RiniTuninettiDevroye_Allerton,RiniTuninettiDevroye}.

In this paper, we consider the effect of cognition on the degrees of freedom (DoF) of a larger network. The considered network consists of a multiple access channel (MAC) and a point-to-point (P2P) link. This setup was studied in \cite{ChaabanSezgin_PIMAC} where it was called the (PIMAC) (the partial version of the IMAC \cite{ChaabanSezginBandemerPaulraj}), but in this paper we deal with the cognitive variant of the PIMAC which we call the (cPIMAC). The cPIMAC with a cognitive P2P transmitter was previously studied in \cite{BaruchSridharanVishwanathVerduShamai} where its capacity with weak interference was derived. Here, we study the DoF of the setup while considering different cases of cognition. Namely, we consider the cases where: (1) none of the  transmitters is cognitive, (2) the P2P transmitter is cognitive, (3) one of the MAC transmitters is cognitive, (4) both MAC transmitters are cognitive. We obtain DoF upper bounds for each of the aforementioned cases. The simple scheme of time division multiplexing is sufficient for achieving the DoF upper bound in cases (1) and (2) equal to 1 DoF. The other cases (3) and (4) have 3/2 DoF achievable by using interference alignment in the complex plane and interference neutralization. Interestingly, there is no difference if one of the MAC transmitters or both of them are cognitive from a DoF point of view. As a consequence, the signaling overhead required for providing side information to the cognitive user is reduced in comparison to having two cognitive transmitters.

We introduce the system model in section \ref{Model}. We give the main result of the paper in section \ref{MainResult}. The upper and lower bounds are derived in sections \ref{UB} and \ref{LB} respectively, and we conclude with section \ref{Conclusion}. We use $C(x)$ to denote $\log(1+x)$.

\section{System Model}
\label{Model}

\begin{figure}[t]
\centering
\includegraphics[width=0.8\columnwidth]{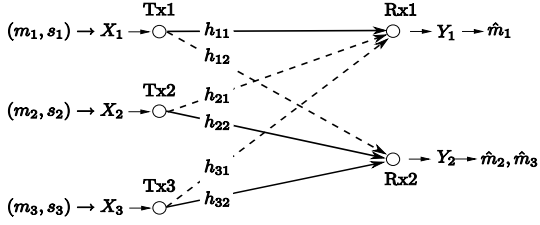}
\caption{The cPIMAC system model.}
\label{Fig:cpIMAC}
\end{figure}

Consider a transmitter Tx1 using a point-to-point (P2P) channel to communicate with receiver Rx1. Two other transmitters Tx2 and Tx3 communicate with receiver Rx2, thereby forming a multiple access channel (MAC), using the same communications medium as the pair (Tx1, Rx1). The transmit messages of transmitters Tx1, Tx2 and Tx3 are $m_1$,$m_2$, and $m_3$ respectively. The first receiver Rx1 wants to decode $m_1$ and the second receiver Rx2 wants to decode $m_2$ and $m_3$. We assume that transmitter $k$ has in addition to message $m_k$ another message $s_k$ with $s_k\in\{m_1,m_2,m_3,(m_2,m_3)\}$, where $s_k\neq m_k$ (to be specified later in this section). Thus, the messages available at transmitter $k$ are $(m_k,s_k)$ as shown in Figure \ref{Fig:cpIMAC}. We call the resulting setup a cognitive partial interference MAC (cpIMAC). 

The purpose of using message $s_k$ is to allow different cognition combinations in the given cpIMAC. We can consider the pair (Tx1, Rx1) to be the primary system and the MAC from Tx2 and Tx3 to Rx2 as the secondary one, or vice versa. Hence, we can distinguish between four different cases:
\begin{itemize}
\item{Case 1:} None of the transmitters is cognitive, i.e. $s_k=\phi\ \forall k\in\{1,2,3\}$;
\item{Case 2:} Only transmitter Tx1 is cognitive, i.e. $s_1\in\{m_2,m_3,(m_2,m_3)\}$ and $s_2,s_3=\phi$;
\item{Case 3:} Only transmitter Tx2 is cognitive, i.e. $s_2=m_1$ and $s_1,s_3=\phi$, or only transmitter Tx3 is cognitive, i.e. $s_3=m_1$ and $s_1,s_2=\phi$; or
\item{Case 4:} Both transmitters Tx2 and Tx3 are cognitive, i.e. $s_2=s_3=m_1$ and $s_1=\phi$.
\end{itemize}

At time instant $i$, transmitter $k$ transmits the symbol $X_{k,i}$ where $k\in\{1,2,3\}$. The corresponding received signals at receivers $j\in\{1,2\}$ can be written as
\begin{align*}
Y_{j,i}&=\sum_{k=1}^3h_{kj}X_{k,i}+Z_{j,i},
\end{align*}
where $h_{kj}\in\mathbb{C}$ denotes the channel coefficient (cf. Figure \ref{Fig:cpIMAC}), and $Z_{j,i}$ is a realization of an i.i.d. circularly symmetric complex noise $Z_j$ with $Z_j\sim\mathcal{CN}(0,1)$. The transmitters have a power constraint $\mathbb{E}[|X_{k,i}|^2]\leq P$. The message $m_k$ is chosen independently from a set $\mathcal{M}_k=\{1,\dots,2^{nR_k}\}$. The messages available at each transmitter $k$, $(m_k,s_k)$, are encoded into length $n$ codewords $X_{k}^n$ using encoding functions $X_{k}^n=f_{k}(m_{k},s_k)$. Receivers Rx1 and Rx2 decode $\hat{m}_{1}$ and $(\hat{m}_2,\hat{m}_{3})$ by using decoding functions $g_1(Y_1^n)$ and $g_2(Y_2^n)$, respectively. This procedure induces an error probability whose average over all messages is denoted $P_e$. An $(n,2^{nR_{1}},2^{nR_{2}},2^{nR_3},P_e)$ code for the cPIMAC consists of encoding functions, decoding functions, and message sets, with decoding error probability $P_e$. A rate triple $(R_{1},R_{2},R_3)$ is said to be achievable if there exists an $(n,2^{nR_{1}},2^{nR_{2}},2^{nR_3},P_e)$ code such that $P_e$ can be made arbitrarily small by increasing $n$. The set of all achievable rate triples is the capacity region of the cpIMAC denoted $\mathcal{C}$. The total degrees of freedom (DoF) of this setup, denoted $d_\Sigma$, is defined as $d_\Sigma=\lim_{P\to\infty}\frac{C_\Sigma(P)}{\log(P)}$, where for a given power constraint $P$, $C_\Sigma(P)=\max_{(R_1,R_2,R_3)\in\mathcal{C}}R_\Sigma$, and $R_\Sigma=R_1+R_2+R_3$. We will denote the DoF $d_\Sigma$ in the 4 cases listed above by $d_\Sigma^{[l]}$ where $l\in\{1,\dots,4\}$ and $l$ refers to cases 1 to 4 listed above. 

\section{Main result}
\label{MainResult}
By studying the DoF of the cpIMAC, we obtain the main result of this paper which is stated in the following theorem.
\begin{theorem}
\label{Theorem:DoF}
The DoF $d_\Sigma$ of the cpIMAC in its 4 variants given in Section \ref{Model} are given as follows
\begin{align*}
d_\Sigma^{[1]}=d_\Sigma^{[2]}=1,\quad\text{and}\quad d_\Sigma^{[3]}=d_\Sigma^{[4]}=\frac{3}{2}.
\end{align*}
\end{theorem}

The achievability and converse proofs of this theorem are given in the following sections. Notice that while allowing Tx1 to be cognitive (case (2)) does not provide any DoF gain compared to the non cognitive case (1), interestingly, making any of Tx2 or Tx3 cognitive increases the DoF from 1 to 3/2 as seen in cases (3) and (4). A cognitive MAC transmitter is thus (asymptotically) more valuable than a cognitive P2P transmitter in a cPIMAC.

\section{Upper Bounds}
\label{UB}
In this section, we develop the upper bounds necessary for obtaining the results in Theorem \ref{Theorem:DoF}. We start by stating the following bounds from Fano's inequality for $k\in\{1,2,3\}$
\begin{align}
\label{nR3}
nR_k\leq I(m_k;Y_k^n)&+n\varepsilon_{kn},\quad \varepsilon_{kn}\to0\text{ as }n\to\infty.
\end{align}

\subsection{Case (1):}
\label{Case1}
Now we consider case (1), the non-cognitive case. The rate $nR_3$ in \eqref{nR3} can be further upper bounded by giving $m_2$ as side information. Thus we can write (with $\varepsilon_{n}=\varepsilon_{1n}+\varepsilon_{2n}+\varepsilon_{3n}$)
\begin{align}
n(R_\Sigma-\varepsilon_{n})&\leq I(m_1;Y_1^n)+I(m_2;Y_2^n)+I(m_3;Y_2^n,m_2)\nonumber\\
&\stackrel{(a)}{=} I(m_1;Y_1^n)+I(m_2;Y_2^n)+I(m_3;Y_2^n|m_2)\nonumber\\
&\stackrel{(b)}{=} I(m_1;Y_1^n)+I(m_2,m_3;Y_2^n)\nonumber\\
&\leq I(m_1;Y_1^n,Y_2^n,m_2,m_3)+I(m_2,m_3;Y_2^n)\nonumber\\
&\stackrel{(c)}{=} I(m_1;Y_2^n|m_2,m_3)+I(m_1;Y_1^n|Y_2^n,m_2,m_3)\nonumber\\
&\quad+I(m_2,m_3;Y_2^n)\nonumber\\
\label{UB1}
&\stackrel{(d)}{=} I(\vec{m};Y_2^n)+I(m_1;Y_1^n|Y_2^n,m_2,m_3),
\end{align}
where $\vec{m}=(m_1,m_2,m_3)$, $(a)$ follows from the independence of $m_2$ and $m_3$, $(b)$ from the chain rule of mutual information, $(c)$ from the independence of the messages, and $(d)$ from the chain rule. The first term in (\ref{UB1}) can be bounded as follows
\begin{align*}
I(\vec{m};Y_2^n)&=h(Y_2^n)-h(Y_2^n|\vec{m})\nonumber\\
&\stackrel{(e)}{=}\sum_{i=1}^n\left[h(Y_{2i}|Y_2^{i-1})-h(Y_{2i}|Y_{2}^{i-1},X_1^n,X_2^n,X_3^n)\right]\nonumber\\
&\stackrel{(f)}{\leq}\sum_{i=1}^n\left[h(Y_{2i})-h(Z_{2i}|Z_{2}^{i-1})\right]\nonumber\\
&\stackrel{(g)}{=}\sum_{i=1}^nh(Y_{2i})-n\log(\pi e)\nonumber\\
&\stackrel{(h)}{\leq} nC(h_{12}^2P+h_{22}^2P+h_{32}^2P)
\end{align*}
where $(e)$ follows using the Markov chain $\vec{m}\to (X_1^n,X_2^n,X_3^n)\to Y_2^n$, $(f)$ since conditioning does not increase entropy, $(g)$ since the noise is i.i.d. Gaussian, and $(h)$ since the circularly symmetric complex Gaussian distribution maximizes the differential entropy under a covariance constraint. The second term in (\ref{UB1}) can be rewritten as follows
\begin{align}
&I(m_1;Y_1^n|Y_2^n,m_2,m_3)\nonumber\\
&=\sum_{i=1}^n\left[h(Y_{1i}|Y_2^n,Y_1^{i-1},m_2,m_3)-h(Y_{1i}|Y_2^n,Y_1^{i-1},\vec{m})\right]\nonumber\\
\label{UB3}
&\stackrel{(i)}{\leq}\sum_{i=1}^n\left[h(Y_{1i}|Y_{2i},X_{2i},X_{3i})-h(Y_{1i}|Y_2^n,Y_1^{i-1},X_2^n,X_3^n,X_1^n)\right]\nonumber
\end{align}
where $(i)$ follows since: in case (1), $X_k^n=f_k(m_k)$, $k\in\{1,2,3\}$, conditioning reduces entropy, and from the Markov chains $(m_2,m_3)\to (X_{2i},X_{3i})\to Y_{1i}$ and $\vec{m}\to (X_1^n,X_2^n,X_3^n)\to (Y_1^n,Y_2^n)$. We proceed as follows
\begin{align}
&I(m_1;Y_1^n|Y_2^n,m_2,m_3)\nonumber\\
&\leq\sum_{i=1}^n\left[h(h_{11}X_{1i}+Z_{1i}|h_{12}X_{1i}+Z_{2,i})-h(Z_{1i}|Z_{2,i},Z_1^{i-1})\right]\nonumber\\
&\stackrel{(j)}{=}\sum_{i=1}^nh(h_{11}X_{1i}+Z_{1i}|h_{12}X_{1i}+Z_{2,i})-n\log(\pi e)\nonumber\\
&\stackrel{(k)}{\leq}nC\left(\frac{h_{11}^2P}{1+h_{12}^2P}\right),
\end{align}
where $(j)$ follows since $Z_{1i}$ is i.i.d. Gaussian, and $(k)$ since the circularly symmetric complex Gaussian distribution maximizes the conditional differential entropy under a covariance constraint \cite{Thomas}. Combining terms and letting $n\to\infty$ yields
\begin{align*}
R_\Sigma&\leq C(P(h_{12}^2+h_{22}^2+h_{32}^2))+C\left(\frac{h_{11}^2P}{1+h_{12}^2P}\right)\\
&=\log(P)+o(\log(P)) \Rightarrow d_\Sigma^{[1]}\leq 1.
\end{align*}

\subsection{Case (2):}
The upper bound for case (2) follows using similar steps as case (1). In fact, some differences exist but these differences do not affect the DoF. For instance, if Tx1 knows $m_2$, then $X_1$ can be correlated with $X_2$. Similarly if it knows $m_3$ or both $m_2$ and $m_3$. In general, we can have correlation coefficients $\rho_{12}$ between $X_1$ and $X_2$ and $\rho_{13}$ between $X_1$ and $X_3$. These correlations represent all 3 cases of $s_1\in\{m_2,m_3,(m_2,m_3)\}$, since if Tx1 does not know $m_2$ e.g. then $\rho_{12}=0$. By taking this correlation coefficient into account, we get $R_\Sigma\leq\log(P)+o(\log(P))$ which is again equivalent to $d_\Sigma^{[2]}\leq1$.

\subsection{Case (3):}
Assume that Tx2 knows $m_1$, or in other words $s_2=m_1$. Notice that in this case, the step taken in (\ref{UB3}) fails since $X_2^n=f_2(m_2)$ does not hold anymore. In fact, $X_2^n$ is now a function of both $m_1$ and $m_2$. However, taking this into account and proceeding as in section \ref{Case1} we obtain a DoF upper bound of 2: $d_\Sigma^{[3]}\leq2$. But this DoF bound is not tight since as we show next, we can use a different approach to obtain an upper bound $d_\Sigma^{[3]}$ of 3/2. For this purpose, we have
\begin{align}
&n(R_1+R_2-\varepsilon_{1n}-\varepsilon_{2n})\nonumber\\
&\leq I(m_1;Y_1^n)+I(m_2;Y_2^n)\nonumber\\
&\leq I(m_1;Y_1^n,m_3)+I(m_2;Y_2^n,Y_1^n,m_1,m_3)\nonumber\\
&\stackrel{(a)}{\leq} I(m_1;Y_1^n|m_3)+(m_2;Y_1^n|m_1,m_3)\nonumber\\
&\quad+I(m_2;Y_2^n|Y_1^n,m_1,m_3)\nonumber\\
\label{MTB12}
&\stackrel{(b)}{\leq} I(m_1,m_2;Y_1^n|m_3)+I(m_2;Y_2^n|Y_1^n,m_1,m_3)
\end{align}
where $(a)$ follows from the independence of the messages and the chain rule and $(b)$ follows from the chain rule. Now, using similar arguments like those used for case (1), we obtain
\begin{align*}
I(m_1,m_2;Y_1^n|m_3)\leq nC(P(h_{11}^2+h_{21}^2+2h_{11}h_{21}\Re[\rho_{12}]))
\end{align*}
where $\rho_{12}=\mathbb{E}[X_1^*X_2]/P$ is the correlation coefficient between $X_1$ and $X_2$ which are in this case correlated since $X_2^n=f_2(m_2,m_1)$. For the second term in \eqref{MTB12} we have
\begin{align*}
I(m_2;Y_2^n|Y_1^n,m_1,m_3)\leq nC\left(\frac{h_{22}^2P}{1+h_{21}^2P}\right).
\end{align*}
Thus
\begin{align}
\label{UB4}
R_1+R_2&\leq \log(P)+o(\log(P)).
\end{align}
Similarly, we can obtain 
\begin{align}
\label{UB5}
R_1+R_3\leq \log(P)+o(\log(P))
\end{align}
and
\begin{align}
\label{UB6}
R_2+R_3\leq \log(P)+o(\log(P)).
\end{align}
Adding up (\ref{UB4}), (\ref{UB5}) and (\ref{UB6}), we get $R_\Sigma\leq \frac{3}{2}\log(P)+o(\log(P))$, which leads to the desired upper bound $d_\Sigma^{[3]}\leq\frac{3}{2}$.

\subsection{Case (4):}
If both Tx2 and Tx3 know $m_1$, i.e. $s_2=s_3=m_1$, then similar to case (3) we can show that $d_\Sigma^{[4]}\leq\frac{3}{2}$. Now that we have obtained DoF upper bounds for the cpIMAC, we can proceed to establish the achievability of these upper bounds.

\section{Achievability}
\label{LB}
The first and the second cases have the same DoF upper bound, and hence they also have the same DoF achieving scheme. We start by considering these two cases and give their DoF achieving scheme.

\subsection{Cases (1) and (2):}
In both cases, the following DoF upper bound holds $d_\Sigma^{[1]}\leq1$. But this upper bound is achievable using simple schemes, like time division multiplexing or decoding all signals at both receivers, which achieve
\begin{align*}
R _\Sigma&\leq\frac{1}{2}C(2h_{11}^2P)+\frac{1}{2}C(2P(h_{22}^2+h_{32}^2)),\\
R_\Sigma&\leq C(P\min\{h_{11}^2+h_{21}^2+h_{31}^2,h_{12}^2+h_{22}^2+h_{32}^2\})\nonumber,
\end{align*}
respectively, each of which achieves 1 DoF.

\subsection{Case (3):}
The achievability of 3/2 DoF in this case is guaranteed by using interference alignment and interference neutralization. A combination of interference alignment and neutralization in a different network (2x2x2 IC) was also recently studied in \cite{GouJafarJeonChung}. We consider the case where Tx2 is cognitive, the case when Tx3 is cognitive follows similarly. 

Let us encode the messages $(m_k,s_k)$ at transmitter $k$ to a real valued codeword $x_k^n$, i.e. $x_{ki}\in\mathbb{R}$, $i\in\{1,\dots,n\}$. We will drop the time index for simplicity and use $x_k$ instead.  Let the complex valued transmit symbols be denoted $X_k\in\mathbb{C}$ which will be constructed from $x_k$ as we explain next. The complex valued symbols $X_k$ can be expressed as 2 a dimensional real vector $\vec{X}_k\in\mathbb{R}^2$ as follows
\begin{align*}
\vec{X}_k=\left[\begin{array}{c}\Re[X_k]\\\Im[X_k]\end{array}\right].
\end{align*}
Using this notation, let us construct $\vec{X}_k$ by using
\begin{align*}
\vec{X}_1=\vec{V}_1x_1,\ \ \ \ \vec{X}_2=\vec{V}_2x_2+\vec{V}_0x_1,\ \ \ \  \vec{X}_3=\vec{V}_3x_3,
\end{align*} 
where $\vec{V}_0,\vec{V}_1,\vec{V}_2$, and $\vec{V}_3$ are $2\times1$ real valued precoding vectors. The complex valued channel coefficients $h_{kj}=|h_{kj}|e^{j\phi_{kj}}$ can be also expressed as $2\times2$ real matrices as follows \cite{CadambeJafarWang}
\begin{align*}
h_{kj}=|h_{kj}|
\left[\begin{array}{cc}\cos{\phi_{kj}} & -\sin{\phi_{kj}}\\
\sin{\phi_{kj}} & \cos{\phi_{kj}}\end{array}\right]=|h_{kj}|\mat{U}_{kj}.
\end{align*}

Thus, the received signal $Y_j\in\mathbb{C}$, expressed in its equivalent 2 dimensional real representation $\vec{Y}_j$ becomes
\begin{align*}
\vec{Y}_j&=\left[\begin{array}{c}\Re[Y_j]\\\Im[Y_j]\end{array}\right]\\
&=(|h_{1j}|\mat{U}_{1j}\vec{V}_1+|h_{2j}|\mat{U}_{2j}\vec{V}_0)x_1+|h_{2j}|\mat{U}_{2j}\vec{V}_2x_2\nonumber\\
&\quad+|h_{3j}|\mat{U}_{3j}\vec{V}_{3}x_3+\vec{Z}_j.
\end{align*}
If we design $\vec{V}_k$ such that
\begin{align*}
\mat{U}_{21}\vec{V}_2=\mat{U}_{31}\vec{V}_3,\quad|h_{12}|\mat{U}_{12}\vec{V}_1=-|h_{22}|\mat{U}_{22}\vec{V}_0
\end{align*}
then, we align interference at Rx1 and we neutralize interference at Rx2. This is simply accomplished by choosing $\vec{V}_3$ and $\vec{V}_0$ at random, and then choosing
\begin{align*}
\vec{V}_2=\mat{U}_{21}^{-1}\mat{U}_{31}\vec{V}_3,\quad \vec{V}_1=-\frac{|h_{22}|}{|h_{12}|}\mat{U}_{12}^{-1}\mat{U}_{22}\vec{V}_0
\end{align*}
Using this precoding vector construction, we get
\begin{align*}
\vec{Y}_1&=\left(\frac{-|h_{11}||h_{22}|}{|h_{12}|}\mat{U}_{11}\mat{U}_{12}^{-1}\mat{U}_{22}+|h_{21}|\mat{U}_{21}\right)\vec{V}_0x_1\nonumber\\
&\quad+\mat{U}_{31}\vec{V}_3(|h_{21}|x_2+|h_{31}|x_3)+\vec{Z}_1\\
&=\widetilde{\mat{U}}_{11}\vec{V}_0x_1+\mat{U}_{31}\vec{V}_3(|h_{21}|x_2+|h_{31}|x_3)+\vec{Z}_1\\
\vec{Y}_2&=|h_{22}|\mat{U}_{22}\mat{U}_{21}^{-1}\mat{U}_{31}\vec{V}_3x_2+|h_{32}|\mat{U}_{32}\vec{V}_3x_3+\vec{Z}_2\\
&=\widetilde{\mat{U}}_{22}\vec{V}_3x_2+\widetilde{\mat{U}}_{32}\vec{V}_3x_3+\vec{Z}_2.
\end{align*}
The random choice of $\vec{V}_0$ and $\vec{V}_3$ suffices to ensure the linear independence of $\widetilde{\mat{U}}_{11}\vec{V}_0$ and $\mat{U}_{31}\vec{V}_3$ at Rx1, and the linear independence of $\widetilde{\mat{U}}_{22}\vec{V}_3$ and $\widetilde{\mat{U}}_{32}\vec{V}_3$ at Rx2 is insured by the randomness of the channels. Now receiver Rx1 projects its received signal $\vec{Y}_1$ to the null space of $\mat{U}_{31}\vec{V}_3$ thus zero-forcing interference, and then decodes the real signal $x_1$ interference free. The second receiver can resolve both $x_2$ and $x_3$ from its two dimensional receive space. Each user thus gets 1/2 DoF and as a result, the achievable DoF is 3/2 which is equal to the DoF upper bound for this case.

\subsection{Case (4):}
The achievability of the DoF upper bound in case 4 is exactly the same as case 3. Interestingly, although both transmitters $U_2$ and $U_3$ are cognitive, the side information at one transmitter can be ignored without any impact on the achievable DoF. In other words, in the cpIMAC it is enough to have only one cognitive user from $\{U_2,U_3\}$ in order to achieve 3/2 DoF.

\section{Conclusion}
\label{Conclusion}
We have studied a network with two components, a multiple access channel and a point-to-point channel, where some transmitters are allowed to be cognitive. The resulting setup, the cPIMAC is studied in different variants, and the following results were obtained. If all transmitters are non-cognitive, then the network has 1 DoF. If the point-to-point transmitter is cognitive, the DoF of the network remains 1. However if either, or both the MAC transmitters are cognitive, then the DoF of the network is increases to 3/2 achievable by using a combination of interference neutralization and alignment.

\bibliography{myBib}

\begin{appendices}

\end{appendices}
\end{document}